\documentstyle[12pt,a4,epsf]{article}

\newcommand{\rbibitem}[1]{\bibitem{#1}}
\newcommand{\be}{\begin{equation}}
\newcommand{\ee}{\end{equation}}
\newcommand{\ba}{\begin{eqnarray}}
\newcommand{\ea}{\end{eqnarray}}

\renewcommand{\mathrm}[1]{{\rm #1}}

\begin{document}
\begin{titlepage}
\begin{flushright}
{FTUV/97-53}\\
{IFIC/97-84}\\
\end{flushright}
\vspace{2cm}
\begin{center}
{\large\bf 
Study of the resummation of chiral logarithms in the exponentiated expression 
for the pion form factor \footnote{Work supported in part by CYCIT, 
Spain, under Grant No. AEN-96/1718}} \\

\vfill
{\bf Francisco Guerrero}\\[0.5cm]
 Departament de
 F\'{\i}sica Te\`orica, Universitat de Val\`encia and\\
 IFIC, CSIC - Universitat de Val\`encia,
 C/ del  Dr. Moliner 50, \\ E-46100 Burjassot (Val\`encia),
Spain\\[0.5cm]
\end{center}
\vfill
\begin{abstract}
From the properties of analyticity and unitarity it has been recently 
obtained an exponentiated expression for the pion form factor. In this work 
I show the validity of this expression comparing its order $p^6$ term with 
the one exactly calculated in ChPT.
\end{abstract}
\vspace*{1cm}
\vfill
November 1997
\end{titlepage}

\vspace{0.5cm}
\noindent{\bf 1. \, Introduction}
\vspace{0.5cm}

ChPT allows us to compute hadronic matrix elements at low energy, where QCD is 
not perturbative. The corresponding results are obtained organized 
as a power expansion in momenta and masses in a perturbative series, 
but the predictions are valid only near the threshold. Different ways of 
summing the series have been developed in order to extend the 
range of validity to higher energies.

In particular, this occurs in the case of the vectorial pion form factor. At 
the moment there at two calculations to the order $p^6$ (i.e. until the 2-loop 
contribution): one numerical \cite{gasmeiss} and one analytical 
\cite{colangelo}.
With this results is possible to explain the 
experimental data from threshold until some 400 MeV in energy, where the 
effects of the $\rho$ resonance are already visible.
To raise this maximum energy is necessary to include the $\rho$ resonance 
explicitly and to resum the series.

Being based in the properties of analyticity and unitarity some resummations 
exist in the literature. For instance, the Gounaris-Sakurai parametrization
\cite{gousak} starts from an extended effective range formula and imposing 
certain conditions on the phase shift obtains a propagator-like expression 
for the form factor that resums all the orders.  Another example is that of 
the inverse amplitude method. It applies a dispersion relation to $F^{-1}$, 
calculated with ChPT,
instead of applying it to $F$, the form factor. It works better 
because the imaginary part of $F^{-1}$ in the elastic case is a better 
approximation to data than just with $F$ \cite{truong,hannah1,hannah2}.

In this letter I will deal with the exponentiated parametrization obtained 
in \cite{guerrpich}, where using Watson theorem \cite{watson}, 
Vector Meson Dominance (VMD) and 
the Omn\`es equation \cite{omnes} we were able to resum the contribution 
of the final state  
interaction of the pions 
into an exponential multiplied by the $\rho$ propagator.

To obtain that parametrization we started with the pion form factor calculated 
at order $p^4$ in ChPT plus the contribution from the $\rho$ exchange, obtained 
with an effective chiral theory including the resonances of the lightest 
vectorial octet. After using Vector Meson Dominance and avoiding to include 
twice the order $p^4$ contribution from the local terms in both ChPT 
calculation (i.e. the $L_9$ constant) and the $\rho$ propagator we got the 
expression (restricted to two flavours)

\be
F(s)=\frac{M^2_{\rho}}{M^2_{\rho}-s}-\frac{s}{96\pi^2 f^2_{\pi}} 
A(m^2_{\pi}/s, m^2_{\pi}/M^2_{\rho})
\ee

\noindent
where
\be      \label{hola}
A(m^2_{\pi}/s, m^2_{\pi}/M^2_{\rho})=\ln(m^2_{\pi}/M^2_{\rho})+
\frac{8m^2_{\pi}}{s}-
\frac{5}{3}+\sigma^3 \ln \left( \frac{\sigma+1}{\sigma-1} \right)
\ee

\noindent
with
\be
\sigma=\sqrt{1-4m^2_{\pi}/s}
\ee

\noindent
The first term sums the local terms to all orders whereas the second term 
gives the contribution of the final state interaction of the pions 
to order $p^4$. 
The next step was to sum that pion interaction to all orders too. Using the 
Watson theorem \cite{watson} we obtained the imaginary part of the form 
factor and using the Omn\`es equation \cite{omnes} we were able to obtain an 
exponentiated expression for the pion form factor. The needed phase shift used 
in the Omn\`es solution is the one coming from the $\pi\pi$ scattering at 
tree level in ChPT.

Matching the Omn\`es solution with the equation (\ref{hola}) we got the 
following expression

\be     \label{caracola}
F(s)=\frac{M^2_{\rho}}{M^2_{\rho}-s} \exp \, \left\{ \frac{-s}{96\pi^2 
f^2} A(m^2_{\pi}/s,m^2_{\pi}/M^2_{\rho}) \right\}
\ee

\noindent
Finally we included the $\rho$ width in the parametrization. 
Calculating with the effective chiral theory with resonances mentioned above 
and making a Dyson summation we saw that the result obtained 
was equivalent to shift 
the imaginary part from the exponent in equation (\ref{caracola}) 
to the denominator in the propagator. 
Making the shift we got our final result for the parametrization,

\be    \label{capullo}
F(s)=\frac{M^2_{\rho}}{M^2_{\rho}-s-iM_{\rho} \Gamma_{\rho}(s)} 
\exp \, \left\{ \frac{-s}{96\pi^2 
f^2} \hbox{Re} \, A(m^2_{\pi}/s,m^2_{\pi}/M^2_{\rho}) \right\}
\ee

\noindent
where

\be
\Gamma_{\rho} (s)=\frac{M_{\rho} s}{96\pi f^2_{\pi}} \sigma^3 \theta 
(s-4m^2_{\pi})
\ee

\noindent
This parametrization fits the experimental data perfectly up to 1 GeV
for both the modulus squared and the phase shift of the pion form factor.
The 
final expression has two different parts: the $\rho$ propagator, which sums 
the local 
terms to all orders 
in the low energy chiral expansion, and the exponential 
which sums the final state interaction between the two pions.

In this work I will check the usefulness of this parametrization. 
Since we started from the order $p^4$ form factor to obtain the exponential
I will compare the $p^6$ contribution predicted by the exponential with the 
exact calculation in ChPT \cite{colangelo}.

\vspace{0.5cm}
\noindent{\bf 2. \, Order $p^6$ term in ChPT}
\vspace{0.5cm}

In the exact and analytical result \cite{colangelo}
of the vectorial pion form factor in ChPT with $SU(2)_L \times SU(2)_R$ the 
expression given by the authors is defined with the help of some functions.

Here I modify slightly the presentation in order to show clearer the 
comparison with the exponentiated parametrization. I will write only the $p^6$ 
term, that I will denote by $F^{(6)}_{ChPT} (s)$, as an expansion in powers of 
the logarithm

\be
L(s)=\ln \frac{1+\sigma}{1-\sigma}
\ee

\noindent
for values of $s$ above the $2\pi$ threshold.

I will have an expression of the following form 

\be
F^{(6)}_{ChPT} (s)=a_0 + a_1 L(s) + a_2 (L(s))^2 + a_3 (L(s))^3
\ee

\noindent
It has to be remembered that the 1-loop calculation only contributes to the 
first two terms $a_0$ and $a_1$, therefore $a_2$ and $a_3$ are strictly 
coming from order $p^6$.

Taking the result from \cite{colangelo} the functions $a_i$ are, at 2-loop order, 
the following ones

\begin{eqnarray}
&a_0 &= \left[ \frac{sm^2_{\pi}}{6(16\pi^2 f^2)^2} \overline f_1 + 
\frac{s^2}{(16\pi^2 f^2)^2} \overline f_2 + \left( \frac{m^2_{\pi}}{16\pi^2 
f^2} \right)^2 \left\{ \left[ 
\overline{\ell_2}-\overline{\ell_1}+\frac{\overline{\ell_6}}{2}-\frac{3}{2} 
\frac{\overline{\ell_3}}{x} \right] \frac{x^2}{27} 
(1+3\sigma^2)\right. \right. \nonumber \\
&    &-\frac{x^2}{30} \overline{\ell_4} + \frac{3191}{6480} x^2 +
\frac{223}{216} x - \frac{16}{9} - \frac{\pi^2 x}{540} (37 x +15) + 
\frac{1}{54} (7x^2-151x+99)  \nonumber \\
&    &\left. \left. -\frac{\pi^2}{72 x} (x^3-30 x^2+78x-128)+
8\pi^2 (x^2-\frac{13}{3} x-2) \left( \frac{1}{192} 
-\frac{1}{32\pi^2}-\frac{1}{48x\sigma^2}\right) \right\} \right]
\nonumber \\
&    &+i \left[\left( \frac{m^2_{\pi}}{16\pi^2 f^2}\right)^2  \left\{ \left( 
\overline{\ell_2}-\overline{\ell_1}+\frac{\overline{\ell_6}}{2}-\frac{3}{2}
\frac{\overline{\ell_3}}{x} \right) \frac{\pi x^2 \sigma^3}{18} + 
\frac{\pi\sigma}{108} (7x^2-151x+99)\right. \right. \nonumber \\
&    & \left. \left. + 8 \pi^2 (x^2-\frac{13}{3} x-2)  
\frac{1}{16\pi x \sigma} \right\} 
\right]   
\end{eqnarray} 
 
\begin{eqnarray}
& a_1 &= \left( \frac{m^2_{\pi}}{16\pi^2 f^2} \right)^2 \left\{ \left[ 
- \left( \overline{\ell_2}-\overline{\ell_1}+\frac{\overline{\ell_6}}{2}
-\frac{3}{2} \frac{\overline{\ell_3}}{x} \right) \frac{x^2 \sigma^3}{18}
-\frac{\sigma}{108} (7x^2-151x+99)
\right. \right. \nonumber \\
&     & \left. \left.  +8\pi^2 \left(x^2-\frac{13}{3} x-2 \right) 
\frac{2\pi^2-3x\sigma^2}{48\pi^2 x^2 \sigma^3} \right]+i \left[\frac{-\pi}{36x 
\sigma^2} \left( x^3-16x^2+120 x-476 \right. \right. \right. \nonumber \\
&     & \left. \left. \left. +512/x \right) \right] \right\}
\end{eqnarray}

\begin{eqnarray}
& a_2 &= \left( \frac{m^2_{\pi}}{16\pi^2 f^2} \right)^2 \left\{ 
\left[ \frac{1}{72x \sigma^2} \left( x^3-16x^2+120 x-476+512/x \right)\right]
+i\left[\frac{\pi}{2x^2 \sigma^3} \right. \right. \nonumber \\
&     & \left. \left. \left( x^2-\frac{13}{3} x -2 \right) 
\right] \right\} 
\end{eqnarray}

\begin{eqnarray}
& a_3 &= -\left( \frac{m^2_{\pi}}{16\pi^2 f^2} \right)^2 \left( 
x^2-\frac{13}{3} x -2 \right) \frac{1}{6 x^2 \sigma^3}
\end{eqnarray} 
 
\noindent
Where $x=s/m^2_{\pi}$, and $\ell_i$ are the scale-independent coupling 
constants of 
$SU(2)_L \times SU(2)_R$ ChPT at order $p^4$ and $f_1$ and $f_2$ at order 
$p^6$.
As it can be seen $a_3$ is real (its imaginary part 
appears at order $p^8$), whereas $a_0$, $a_1$ and $a_2$ are complex.

Another important characteristic is that these functions are divergent at 
threshold (i. e. $\sigma=0$). Obviously $F^{(6)}_{ChPT} (4m^2_{\pi})$ is not, 
since the divergencies from the functions, combined with the powers on 
$\sigma$ appearing in 
the expansion of $L(s)$ for small $\sigma$, cancel each other.
This divergences are originated in the loops exchanged in the u and 
t-channels. This implies that they are not expected to appear in the 
exponenciated resummation, since the latter sums only the final state 
interaction of the pions in the s-channel.

\vspace{0.5cm}
\noindent{\bf 3. \, Order $p^6$ term in the exponential parametrization}
\vspace{0.5cm}

Expanding in powers of momenta I get the order $p^6$ term  given by the 
equation (\ref{capullo}).  
Now, if I do the same as in the case of the ChPT calculation and I expand in powers 
of the logarithm $L(s)$ in the form

\be
F^{(6)}_{exp} (s)=b_0 + b_1 L(s) + b_2 (L(s))^2 + b_3 (L(s))^3
\ee

\noindent
the functions, denoted this time by $b_i$, are

\begin{eqnarray}
& b_0 &= \left[ \frac{s^2}{M^4_{\rho}}-\frac{\Gamma^2}{M^2_{\rho}}+ 
\frac{1}{2} \frac{1}{(96 \pi^2 f^2)^2} \left( s \ln \left( 
\frac{m^2_{\pi}}{M^2_{\rho}}\right) + 8m^2_{\pi}-\frac{5}{3} s \right)^2 
\right.   \nonumber \\
&     &- \left. \frac{s}{96 \pi^2 f^2 M^2_{\rho}} \left(s \ln \left(
\frac{m^2_{\pi}}{M^2_{\rho}}\right) + 8m^2_{\pi}-\frac{5}{3} s \right) \right]
\nonumber \\
&     & i \left[ \frac{2s^2 \sigma^3}{96 \pi f^2 M^2_{\rho}}- \frac{\pi s 
\sigma^3}{(96 \pi^2 f^2)^2} \left( s \ln \left(
\frac{m^2_{\pi}}{M^2_{\rho}}\right) + 8m^2_{\pi}-\frac{5}{3} s \right)\right]
\end{eqnarray}
 
\be
b_1= \frac{s^2 \sigma^3}{(96 \pi^2 f^2)^2} \left[ \left\{ \ln 
\left(\frac{m^2_{\pi}}{M^2_{\rho}}\right) + 8m^2_{\pi}-\frac{5}{3} s- \frac{96 
\pi^2 f^2}{M^2_{\rho}} \right\} -i \pi \sigma^3 \right]
\ee

\be
b_2=\frac{1}{2} \frac{s^2 \sigma^6}{(96 \pi^2 f^2)^2}
\ee

\be
b_3=0
\ee

As it was expected, the values of the $b_i$ functions are not 
divergent at threshold. Therefore the direct comparison between $a_i$ and 
$b_i$ has no sense for $\sigma=0$. I will compare $F^{(6)}_{exp} 
(4m^2_{\pi})$ with $F^{(6)}_{ChPT}(4m^2_{\pi})$ because now both quantities
are finite.

\vspace{0.5cm}
\noindent{\bf 4. \, Comparison}
\vspace{0.5cm}

Now that I have introduced the necessary formulae I can proceed to the 
comparison of both results.

In the way the functions $a_i$ and $b_i$ are written, that is, with the 
complete analytic expression, no similarity can be seen at first sight between 
them.

To understand better the physics behind these formulae is convenient to go to 
the chiral limit ($m_{\pi} =0$) where the similarities become more evident. In 
this limit the $a_i$ functions are finite at threshold so we can compare them
directly with the $b_i$ functions. 

I begin with the highest power of the logarithm, $a_3$ and $b_3$. Here the 
chiral limit is easy

\be
\hat{a_3}=\hat{b_3}=0
\ee

\noindent
The hat means that the quantity is in the chiral limit.

The coefficients for the second power of the logarithm are also equal in this 
limit.

\be
\hat{a_2}=\hat{b_2}=\frac{1}{72} \left( \frac{s}{16 \pi^2 f^2} \right)^2
\ee

\noindent
This is an important result because it means that in the chiral limit the 
dominant logarithms to order $p^6$ are correctly resummed. This fact does not 
occur in resummations based on the [0,1] Pad\'e approximants like for instance 
those from the inverse amplitude method \cite{truong,hannah1,hannah2} or the 
Gounaris-Sakurai parametrization \cite{gousak}.

In the subleading term (i.e. the one linear in the logarithm)
we have the first differences. 
The values for the functions are

\begin{eqnarray}
& \hat{a_1} &= \left( \frac{s}{16 \pi^2 f^2} \right)^2 \left[ -\left\{
\frac{1}{18}
\left(\overline{\ell_2}-\overline{\ell_1}+\frac{\overline{\ell_6}}{2} 
\right) 
+ \frac{7}{108} \right\}-i \frac{\pi}{36} \right]  \nonumber \\
& \hat{b_1} &=\left( \frac{s}{16 \pi^2 f^2} \right)^2 \left[ \frac{1}{36} 
\left\{ \ln \left( \frac{m^2_{\pi}}{M^2_{\rho}} \right) -\frac{5}{3} - 
\frac{96 \pi^2 f^2}{M^2{\rho}} \right\} -i \frac{\pi}{36} \right]
\end{eqnarray}

\noindent
In order to establish a good comparison I have to manipulate the real part of 
$a_1$, in particular the $\overline{\ell_i}$ constants. In \cite{colangelo} we 
can find how to rewrite them in terms of the usual $\ell^r_i$ from Gasser and 
Leutwyler \cite{gasleut1}, and also in the latter we find how to pass to the 
$SU(3)_L \times SU(3)_R$ constants denoted by $L^r_i$. The equivalence is

\be
\overline{\ell_2}-\overline{\ell_1}+\frac{\overline{\ell_6}}{2}=
96 \pi^2 (2L^r_2-4L^r_1-4L^r_3+2L^r_9) - \frac{1}{2} \ln \left( \frac 
{m^2_{\pi}}{\mu^2} \right)
\ee
 
Applying now Vector Meson Dominance in accordance with \cite{ecker} 
at the scale $\mu^2=M^2_{\rho}$ 
(remember that to derive the exponentiated parametrization 
we used VMD) I obtain 

\be
\overline{\ell_2}-\overline{\ell_1}+\frac{\overline{\ell_6}}{2}= \frac{120 
\pi^2 f^2}{M^2_{\rho}}-\frac{1}{2} \ln \left( \frac{m^2_{\pi}}{M^2_{\rho}} 
\right)
\ee

\noindent
In this way the real part of $\hat{a_1}$ now is, in the VMD approximation,

\be
\hbox{Re} \, \hat{a_1}= \left( \frac{s}{16 \pi^2 f^2} \right)^2 \frac{1}{36} 
\left\{ \ln \left( \frac{m^2_{\pi}}{M^2_{\rho}} \right) - \frac{7}{3}- 
\frac{240 \pi^2 f^2}{M^2_{\rho}} \right\}
\ee

\noindent
The correct logarithm is reproduced, however there is a difference with 
$\hat{b_1}$

\be
\hat{a_1}-\hat{b_1}=\left( \frac{s}{16 \pi^2 f^2} \right)^2 \frac{1}{36} 
\left\{\frac{-2}{3}-\frac{144 \pi^2 f^2}{M^2_{\rho}} \right\}
\ee

\noindent
This difference comes from the contribution due to the 
exchange of one $\rho$ in the t-channel in the interaction between the final 
pions. It has to be remembered that the final state interaction, resummed in 
the exponential, comes here from the tree level phase shift $\delta^1_1 (s)$ 
calculated from $\pi\pi$ scattering. It would be necessary to include the 
1-loop term in $\delta^1_1 (s)$ to obtain that contribution. 

Finally I compare the polymonial terms. The chiral limits for $a_0$ and $b_0$ 
are

\begin{eqnarray}
& \hat{a_0} & =\left( \frac{s}{16 \pi^2 f^2} \right)^2 \left[ \left\{ 
{\overline 
f_2}+ \frac{4}{27} \left(\overline{\ell_2}-\overline{\ell_1}+\frac{
{\overline \ell_6}}{2} \right) - 
\frac{\overline{\ell_4}}{30}+\frac{2411}{6480}-\frac{11\pi^2}{270} \right\}
\right.
\nonumber \\
&           & \left. i \left\{\left(\overline{\ell_2}-\overline{\ell_1}
+\frac{
\overline{\ell_6}}{2} \right) \frac{\pi}{18}+\frac{7\pi}{108} \right\} 
\right]  
\end{eqnarray}

\begin{eqnarray}
& \hat{b_0} &= s^2 \left[ \frac{1}{M^4_{\rho}}-\frac{1}{(96 \pi^2 f^2)^2}+ 
\frac{1}{2(96 \pi^2 f^2)^2} \left( \ln \left( \frac{m^2_{\pi}}{M^2_{\rho}} 
\right)-\frac{5}{3} \right)^2 \right. \nonumber \\
&           & \left. - \frac{1}{96 \pi^2 f^2 M^2_{\rho}} \left( \ln 
\left( \frac{m^2_{\pi}}{M^2_{\rho}}\right)-\frac{5}{3} \right)\right]
\nonumber \\
&           &+i s^2 \left[\frac{1}{48\pi f^2 M^2_{\rho}}- \frac{\pi}{(96 \pi^2 
f^2)^2} \left( \ln \left( \frac{m^2_{\pi}}{M^2_{\rho}}\right)-\frac{5}{3} 
\right)\right]
\end{eqnarray}

\noindent
In ${\hat a_0}$ I have already taken ${\overline f_1}=0$ as it is suggested by 
all the authors. This choice is justified because the 
contribution to the 
electromagnetic radius of the pion from $\overline{f_1}$ is 
negligible, since this radius is 
saturated by the $\rho$ resonance (i.e. the $L_9$ constant)

To establish numerical comparisons I will take the following experimental 
values for the ${\overline \ell_i}$ constants \cite{colangelo}

\begin{eqnarray}
& {\overline \ell_1} &=-1.7 \pm 1.0    \nonumber   \\
& {\overline \ell_2} &=6.1 \pm 0.5     \nonumber   \\
& {\overline \ell_3} &=2.9 \pm 2.4     \nonumber   \\
& {\overline \ell_4} &=4.3 \pm 0.9     \nonumber   \\
& {\overline \ell_6} &=16.5 \pm 1.1    
\end{eqnarray}

\noindent
With these values I obtain

\begin{eqnarray}
& {\hat a_0} &=\left[ (1.16 \pm 0.16+0.53 \, {\overline f_2})+
i (1.59 \pm 0.4) \right] \, s^2    \nonumber  \\
& {\hat b_0} &=(3.84 +1.52 \, i) \, s^2
\end{eqnarray}

\noindent
As we can see the imaginary parts coincide. The real parts are equal for 
${\overline f_2}=5.1 \pm 0.3$. This value agrees with previous estimations. 
Some of these estimations are \cite{colangelo} ${\overline f_2} \sim 4.8$, 
\cite{gousak} ${\overline f_2} = 6.9$, \cite{gasmeiss} ${\overline f_2} 
= 6.6$, \cite{hannah1,hannah2} ${\overline f_2} = 3.7$.

After all this colection of formulae it can be concluded that the exponentiated 
parametrization is, in the chiral limit, a good extrapolation for ChPT at 
higher energies. The prediction made for the order $p^6$ term is basically 
equal to the ChPT result (the order $p^4$ term is exact by 
construction). It is different only in the real part of 
$a_1$ (that is, 
since the functions are complex, in one of six), 
where it was expected 
due to the arguments given above.

Once the underlying physics has been seen, analyzing the chiral limit, we can 
study numerically the complete expressions (without any limit)
for $a_i$ and $b_i$.

We have to remember that $a_i$ is divergent at $\sigma=0$, so in that point 
the comparison between ChPT and the exponential has to be done for the total 
sum at order $p^6$ and not term by term.

The values are

\begin{eqnarray}
&F^{(6)}_{ChPT}(s=4m^2_{\pi})&=0.0227 \pm 0.0009   \nonumber \\
&F^{(6)}_{exp}(s=4m^2_{\pi})&=0.0216
\end{eqnarray}

\noindent
They are completely equivalent. For $F^{(6)}_{ChPT}$ I have taken the constant 
${\overline f_2}$ equal to the value obtained above.

When we increase the energy $\sqrt{s}$ the difference between 
$F^{(6)}_{ChPT}$ and $F^{(6)}_{exp}$ also increases. For $\sqrt{s} \sim 0.7$ 
GeV the real part in the exponentiated expression is only a 15\% larger than 
that of ChPT, reaching 33\% around 1 GeV. In the imaginary part the comparison 
is even better keeping the difference around the 3\% for $\sqrt{s}=1$ GeV.

However at so high energies the expansion in ChPT is 
not valid anymore. We have just to remember that around 0.7 GeV the order 
$p^4$ correction has the same value that the tree level one (order $p^2$), and 
the same happens around 0.8 GeV between the orders $p^4$ and $p^6$.

The numerical study allows us to conclude that the most important piece at 
order $p^6$ is the polynomial, followed by the term linear in the logarithm, 
the quadratic and so on.

One delicate point of the exponentiated parametrization is the shift of the 
imaginary part from the exponent to the propagator \cite{guerrpich}.
If we do not do the shift and keep the imaginary part in the exponent the 
terms with logarithms are not modified at all (i. e. the functions $b_1$,$b_2$ 
and $b_3$ remain the same). However the $b_0$ function changes substantially 
in its imaginary part. In the chiral limit its value passes from $\hbox{Im}\, 
{\hat b_0}=1.52 \, s^2$ to $\hbox{Im} \, {\hat b_0}=0.88 \, s^2$, a difference 
of the 50\%. Thus, higher-order corrections are more efficiently summed doing 
the shift, as expected from the Dyson summation of the $\rho$ 
self-energy.      
 
\vspace{0.5cm}
\noindent{\bf 5. Phase shift $\delta^1_1 (s)$}
\vspace{0.5cm}

There is also another test that we can do with the 
exponentiated parametrization and it has to do with the phase 
shift.

The resummation presented in \cite{guerrpich} 
introduces a prediction for the 
phase shift which, as we 
saw then, fits the data perfectly.

Here I present this graphically reconstructing the pion form factor from the 
obtained phase shift

\be
\delta^1_1 (s)= \arctan \left( \frac{M_{\rho} \Gamma_{\rho} (s)}{M^2_{\rho}-s} 
\right)
\ee

\noindent
and using dispersion relations

\be   \label{disp}
F(s)=\exp \left\{ \sum^{n-1}_{k=1} \,\left[\ln F(0) \right]^{(k)} 
\frac{s^k}{k!} \right\} \exp \left\{ \frac{s^n}{\pi} \int \, \frac{dz}{z^n}
\frac{\delta^1_1 (z)}{z-s} \right\}
\ee

\noindent
with

\be
\left. \left[ \ln \, F(0) \right]^{(k)}= \frac{d^{(k)} \ln \, 
(-i F(s)/2)}{ds^k} 
\right |_{s=0}
\ee

\noindent
Its experimental values are

\begin{eqnarray}
&\left[ \ln \, F(0) \right]^{(1)}&= \frac{1}{6} \left< r^2_V \right> = 1.98 \, 
\hbox{GeV}^{-2}   \nonumber \\
&\left[ \ln \, F(0) \right]^{(2)}&=2c_V-\left(\frac{1}{6} \left< r^2_V 
\right>\right)^2=4.13 \, \hbox{GeV}^{-4}
\end{eqnarray}

\noindent
The result shown in the figure is the numerical solution for the dispersion 
integral, eq. (\ref{disp}), with one, two and three subtractions (respectively 
the lower, the upper and the middle curves).

The fit obtained for the data is pretty good, and improves with the number of 
subtractions. This is due to the fact that increasing the number of 
subtractions the 
contribution from lower energies (the better understood region) becomes more 
and more important. 

This plot shows that the prediction given for the phase shift by the 
exponential parametrization is correct.

\vspace{0.5cm}
\noindent{\bf 6. Conclusion}
\vspace{0.5cm}

In \cite{guerrpich} 
we obtained an exponentiated expression for the pion form factor based on 
its properties of analiticity and unitarity.

We started there from the tree level phase shift $\delta^1_1$ and, applying a 
dispersion relation with subtractions, we obtained the Omn\`es solution in 
form of an exponential. Its expansion in powers of momentum gives the form 
factor to order $p^4$ exactly in ChPT, as it had to be, because the complex 
phase of the form factor is the same as the one of the 
$\pi\pi$ scattering amplitude, 
i.e. $\delta^1_1 (s)$. If the latter is to order $p^2$ then its dispersion 
integral has to give the form factor to order $p^4$.

However, the exponentiated parametrization contains all the orders in powers 
of momenta. In particular it contains the order $p^6$ term. We can take 
advantage of this to compare this predicted term with the one coming from the 
already existing calculation of the pion form factor at two loops in ChPT.
In this work I have shown that comparison to study the validity of the 
exponentiated parametrization. 

In the chiral limit I have proved qualitatively that the resummation 
is correct. 
It reproduces the correct factor for the dominant logarithm and a value 
compatible with those in the literature for ${\overline f_2}$.

Numerically we observe that both contributions (ChPT and exponential) 
at order $p^6$ differ only in a 
few per cent in the real parts and less than one per cent in the imaginary 
ones, for the region of energies where the expansion of ChPT has sense 
($\sqrt{s} < 500$ MeV). This indicates that the parametrization 
contains the most relevant contributions at higher orders. The 
ones not included (loops and resonances in t and u-channels) are not 
quantitatively so important.

The difference observed in the functions $a_1$ and $b_1$ between ChPT and the 
exponential can be corrected introducing $\delta^1_1 (s)$ to order $p^4$. In 
that case the correspondance between $F^{(6)}_{ChPT}$ and $F^{(6)}_{exp}$ 
would be exact.

The study done in this work suggests that the resummation 
obtained to that order would be equally useful, and it would 
give a prediction for the order $p^8$ where
no exact calculation in ChPT has been done. 
Remember that to calculate the pion form factor in ChPT to 
order $p^8$ is not possible for the moment because of its complexity and the 
unknowledgement of the many constants appearing 
in the order $p^6$ and $p^8$ terms in the lagrangian of ChPT.

We can finally conclude from this study that the exponentiated parametrization 
gives a resummation correctly defined that includes the most 
relevant pieces at higher orders. 

Future calculations should work out the resummation with the 
order $p^4$ phase shift.
It would be also worthwhile to try to apply 
this parametrization to other channels.

\vspace{0.8cm}
\noindent
{\bf Acknowledgements}
\vspace{0.5cm}

I want to thank Antonio Pich for useful discussions and reading the manuscript 
and M. Antonia D\'{\i}az for her essential help. 
This work has been supported by an FPI scholarship of the Spanish {\it 
Ministerio de Educaci\'on y Cultura}.

\begin{figure}
\begin{center}
\leavevmode\epsfxsize=12cm\epsfbox{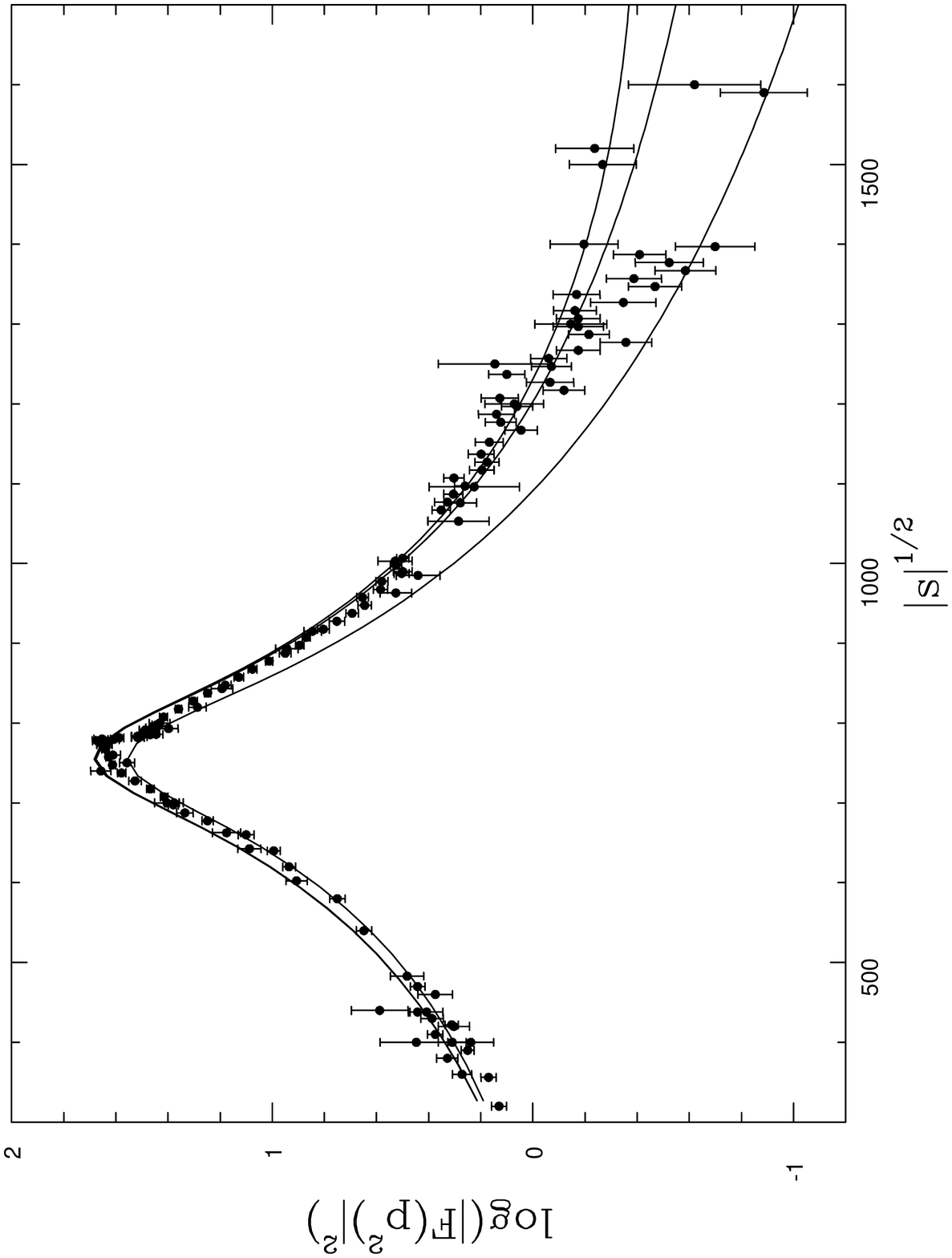}
\end{center}
\end{figure}

\end{document}